\documentclass[letter]{aa} 
\usepackage{graphicx}
\usepackage{txfonts}
\usepackage{natbib}

\newcommand{\tablenotea}[1]{\parbox{15.5cm}{ \indent
\footnotesize{\textsc{Note.--}~#1}}}
\newcommand{\tablenoteb}[1]{\parbox{8.8cm}{ \indent
\footnotesize{\textsc{Note.--}~#1}}}
\newcommand{\tablerefs}[1]{\parbox{18.1cm}{ \indent
\footnotesize{\textsc{References.--}~#1}}}

\newcommand{\nature}{Nature}                   
\newcommand{\ijms}{Int.~J.~Mass Spectr.}       
\begin{document}
\title{Search for anions in molecular sources: C$_4$H$^-$ detection in
L1527\thanks{Based on observations carried out with the IRAM 30m
telescope. IRAM is supported by INSU/CNRS (France), MPG (Germany)
and IGN (Spain).}}
\titlerunning{Search for anions in molecular sources}
\authorrunning{Ag\'undez et al.}

\author{M. Ag\'undez\inst{1}, J. Cernicharo\inst{1}, M. Gu\'elin\inst{2},
M. Gerin\inst{3}, M. C. McCarthy\inst{4} and P. Thaddeus\inst{4}}

\offprints{M. Ag\'undez}

\institute{Departamento de Astrof\'isica Molecular e Infrarroja,
Instituto de Estructura de la Materia, CSIC, Serrano 121, 28006
Madrid, Spain; \email{marce@damir.iem.csic.es,
cerni@damir.iem.csic.es} \and Institut de Radioastronomie
Millim\'etrique, 300 rue de la Piscine, 38406 St. Martin
d'H\'eres, France; \email{guelin@iram.fr} \and LERMA, UMR 8112,
CNRS, Observatoire de Paris and \'Ecole Normale Sup\'erieure, 24
Rue l'Homond, 75231 Paris, France \email{gerin@lra.ens.fr} \and
Harvard-Smithsonian Center for Astrophysics, 60 Garden Street,
Cambridge, MA 02138, USA \email{mccarthy@cfa.harvard.edu,
pthaddeus@cfa.harvard.edu}}

\date{Received ; accepted }


\abstract
{}
{We present the results of a search for the negative ion
C$_4$H$^-$ in various dark clouds, low mass star-forming regions
and photon-dominated regions (PDRs). We have also searched for
C$_6$H$^-$, C$_2$H$^-$ and CN$^-$ in some of the sources.}
{The millimeter-wave observations were carried out with the
IRAM-30m telescope.}
{We detect C$_4$H$^-$, through the J = 9-8 and J = 10-9 rotational
transitions, in the low mass star-forming region L1527. We thus
confirm the tentative detection of the J = 9-8 line recently
reported toward this source. The [C$_4$H$^-$]/[C$_4$H] ratio found
is 0.011 \%, which is slightly lower than the value observed in
IRC +10216, 0.024 \%, but above the 3$\sigma$ upper limit we
derive in TMC-1, $<$ 0.0052 \%. We have also derived an upper
limit for the [C$_6$H$^-$]/[C$_6$H] ratio in the Horsehead Nebula,
and for various anion-to-neutral ratios in the observed sources.
These results are compared with recent chemical models.}
{}


\maketitle
%

\section{Introduction}

We have learnt that negatively charged molecules are present in
the interstellar medium. Following the laboratory detection of the
series of hydrocarbon anions C$_{2n}$H$^-$ (n = 1, 2, 3, 4;
\citealt{mcc06,gup07,bru07a}) and of the smallest member of the
isoelectronic series C$_{2n+1}$N$^-$ (n = 0; \citealt{got07}),
astronomical searches have succeeded in detecting the three
largest anions of the first series in the C-rich envelope around
the evolved star IRC +10216, C$_6$H$^-$ and C$_8$H$^-$ in the dark
cloud TMC-1, and C$_6$H$^-$ toward the low mass star-forming
region L1527 \citep{mcc06,cer07,rem07,bru07b,sak07a}. The smallest
members of each series, C$_2$H$^-$ and CN$^-$, remain undetected
in space.

It has been found that the abundance of the larger anions
C$_6$H$^-$ and C$_8$H$^-$ represents a substantial fraction of
that of their neutral counterparts (a few percent) while the
smaller C$_4$H$^-$ has an abundance much lower than C$_4$H (0.024
\% in IRC +10216). Thus, the anion-to-neutral ratio decreases when
moving from large to small species. This trend was predicted many
years ago by \citet{her81}, who discussed the formation and
potential detectability of molecular anions in interstellar
clouds. It was pointed out that the efficiency of electron
radiative attachment greatly increases for species with a high
electron affinity, such as the radicals C$_{2n}$H and C$_{2n+1}$N,
and with a large number of vibrational states, i.e. a large size.
Thus, the electron radiative attachment rate coefficients seems to
be the crucial parameter controlling the anion-to-neutral ratio.
\citet{mil07} have calculated such rate coefficients and
constructed chemical models which predict anions to be abundant in
dark clouds and PDRs.

To explore how ubiquitous and abundant molecular anions are in the
interstellar medium we have searched for C$_4$H$^-$ with the
IRAM-30m telescope toward various dark clouds, low mass
star-forming regions and PDRs. We have also searched for CN$^-$,
C$_2$H$^-$ and C$_6$H$^-$ in some of the sources. In this Letter
we present the results of these searches, which have led to the
positive detection of C$_4$H$^-$ in the low mass star-forming
region L1527.

\begin{table*}
\caption{Observed line parameters} \label{table-lineparameters}
\centering
\begin{tabular}{l@{ }ll@{}rc@{}c@{}c@{}c@{}c}     
\hline\hline
\multicolumn{1}{l}{}         & \multicolumn{1}{c}{}         & \multicolumn{1}{c}{}           & \multicolumn{1}{c}{Frequency} & \multicolumn{1}{c}{T$_A^*$} & \multicolumn{1}{c}{$\Delta v$}    & \multicolumn{1}{c}{$V_{LSR}$}     & \multicolumn{1}{c}{$\int$T$_A^*$$dv$} & \multicolumn{1}{c}{} \\
\multicolumn{1}{l}{Source}   & \multicolumn{1}{c}{Molecule} & \multicolumn{1}{c}{Transition} & \multicolumn{1}{c}{(MHz)}     & \multicolumn{1}{c}{(mK)}    & \multicolumn{1}{c}{(km s$^{-1}$)} & \multicolumn{1}{c}{(km s$^{-1}$)} & \multicolumn{1}{c}{(mK km s$^{-1}$)} & \multicolumn{1}{c}{$\eta_b$$^c$} \\
\hline
TMC-1$^e$     & C$_4$H       & N=9-8 J=9.5-8.5$^d$                 & 85634.012             & 678(12) & 0.58(8)  & +5.83(4)  & 417(11)  & 0.82 \\
              & C$_4$H       & N=9-8 J=8.5-7.5$^d$                 & 85672.578             & 626(11) & 0.58(8)  & +5.74(4)  & 386(10)  & 0.82 \\
\cline{2-9}
              & C$_4$H$^-$   & J=9-8                               & 83787.263             & 1.5$^a$ & ---      & ---       & 2.8$^b$  & 0.82 \\
              & CN$^-$       & J=2-1                               & 224525.061            & 7.8$^a$ & ---      & ---       & 14.5$^b$ & 0.59 \\
              & C$_2$H$^-$   & J=3-2                               & 249824.940            & 3.5$^a$ & ---      & ---       & 6.6$^b$  & 0.53 \\
\hline
Barnard 1$^e$ & C$_4$H       & N=9-8 J=9.5-8.5$^d$                 & 85634.012             & 145(9)  & 1.21(7)  & +6.79(5)  & 186(8)   & 0.82 \\
              & C$_4$H       & N=9-8 J=8.5-7.5$^d$                 & 85672.578             & 133(8)  & 1.09(7)  & +6.64(5)  & 154(7)   & 0.82 \\
\cline{2-9}
              & C$_4$H$^-$   & J=9-8                               & 83787.263             & 1.2$^a$ & ---      & ---       & 4.4$^b$  & 0.82 \\
              & CN$^-$       & J=2-1                               & 224525.061            & 2.8$^a$ & ---      & ---       & 10.3$^b$ & 0.59 \\
              & C$_2$H$^-$   & J=3-2                               & 249824.940            & 7.9$^a$ & ---      & ---       & 29$^b$   & 0.53 \\
\hline
L134N         & C$_4$H       & N=9-8 J=9.5-8.5                     & 85634.012             & 131(7)  & 0.33(4)  & +2.47(4)  & 46(5)    & 0.82 \\
              & C$_4$H       & N=9-8 J=8.5-7.5                     & 85672.578             & 122(6)  & 0.29(4)  & +2.38(4)  & 38(4)    & 0.82 \\
\cline{2-9}
              & C$_4$H$^-$   & J=9-8                               & 83787.263             & 2.9$^a$ & ---      & ---       & 2.9$^b$  & 0.82 \\
              & CN$^-$       & J=2-1                               & 224525.061            & 14$^a$  & ---      & ---       & 14$^b$   & 0.59 \\
\hline
L1527$^e$     & C$_4$H$^-$   & J=9-8                               & 83787.263             & 13(2)   & 0.62(9)  & +5.80(3)  & 8(1)    & 0.82 \\
              & C$_4$H$^-$   & J=10-9                              & 93096.551             & 11(2)   & 0.59(9)  & +5.90(4)  & 7(1)    & 0.81 \\
              & C$_4$H       & N=9-8 J=9.5-8.5$^d$                 & 85634.012             & 947(11) & 0.74(6)  & +5.95(2)  & 747(11) & 0.82 \\
              & C$_4$H       & N=9-8 J=8.5-7.5$^d$                 & 85672.578             & 871(11) & 0.77(6)  & +5.89(2)  & 712(11) & 0.82 \\
              & C$_4$H       & N=11-10 J=11.5-10.5$^d$             & 104666.566            & 681(10) & 0.75(6)  & +5.92(2)  & 542(10) & 0.79 \\
              & C$_4$H       & N=11-10 J=10.5-9-5$^d$              & 104705.109            & 656(10) & 0.70(6)  & +5.88(2)  & 487(10) & 0.79 \\
              & C$_4$H       & N=12-11 J=12.5-11.5                 & 114182.516            & 712(14) & 0.61(4)  & +5.93(3)  & 462(13) & 0.78 \\
              & C$_4$H       & N=12-11 J=11.5-10.5                 & 114221.039            & 663(13) & 0.58(4)  & +5.85(3)  & 406(12) & 0.78 \\
\cline{2-9}
              & CN$^-$       & J=2-1                               & 224525.061            & 2.3$^a$ & ---      & ---       & 5.5$^b$ & 0.59 \\
              & C$_2$H$^-$   & J=3-2                               & 249824.940            & 3.5$^a$ & ---      & ---       & 6.7$^b$ & 0.53 \\
\hline
L483$^e$      & C$_4$H       & N=9-8 J=9.5-8.5$^d$                 & 85634.012             & 387(15) & 0.60(6)  & +5.27(4)  & 249(12) & 0.82 \\
              & C$_4$H       & N=9-8 J=8.5-7.5$^d$                 & 85672.578             & 332(15) & 0.70(6)  & +5.33(4)  & 249(12) & 0.82 \\
\cline{2-9}
              & C$_4$H$^-$   & J=9-8                               & 83787.263             & 4.5$^a$ & ---      & ---       & 9.3$^b$ & 0.82 \\
              & CN$^-$       & J=2-1                               & 224525.061            & 11.5$^a$ & ---     & ---       & 28$^b$  & 0.59 \\
\hline
Horsehead$^e$ & C$_4$H       & N=9-8 J=9.5-8.5$^d$                 & 85634.012             & 199(10) & 0.86(8)  & +10.72(4) & 181(9)  & 0.82 \\
              & C$_4$H       & N=9-8 J=8.5-7.5$^d$                 & 85672.578             & 155(8)  & 0.81(8)  & +10.69(4) & 133(9)  & 0.82 \\
              & C$_6$H       & $^2\Pi_{3/2}$ J=29.5-28.5 f$^d$     & 81777.898             & 13(2)   & 0.93(16) & +10.76(8) & 13(2)   & 0.82 \\
              & C$_6$H       & $^2\Pi_{3/2}$ J=29.5-28.5 e$^d$     & 81801.254             & 10(2)   & 0.97(21) & +10.81(9) & 11(2)   & 0.82 \\
\cline{2-9}
              & C$_4$H$^-$   & J=9-8                               & 83787.263             & 1.9$^a$ & ---      & ---       & 5.0$^b$ & 0.82 \\
              & C$_6$H$^-$   & J=30-29                             & 82608.285             & 1.5$^a$ & ---      & ---       & 4.6$^b$ & 0.82 \\
              & CN$^-$       & J=2-1                               & 224525.061            & 2.2$^a$ & ---      & ---       & 4.9$^b$ & 0.59 \\
\hline
Orion Bar$^e$ & C$_4$H       & N=9-8 J=9.5-8.5$^d$                 & 85634.012             & 58(5)   & 1.86(11) & +10.77(6) & 116(7)  & 0.82 \\
              & C$_4$H       & N=9-8 J=8.5-7.5$^d$                 & 85672.578             & 55(10)  & 2.54(13) & +10.60(9) & 149(8)  & 0.82 \\
\cline{2-9}
              & C$_4$H$^-$   & J=9-8                               & 83787.263             & 1.2$^a$ & ---      & ---       & 8.4$^b$ & 0.82 \\
              & CN$^-$       & J=2-1                               & 224525.061            & 3.2$^a$ & ---      & ---       & 18$^b$  & 0.59 \\
\hline
NGC 7023$^{e,f}$ & CN$^-$    & J=2-1                               & 224525.061            & 6.4$^a$ & ---      & ---       & 16.4$^b$& 0.59 \\
              & C$_2$H$^-$   & J=3-2                               & 249824.940            & 9.5$^a$ & ---      & ---       & 20$^b$  & 0.53 \\
\hline
\end{tabular}
\tablenotea{The line parameters have been obtained from Gaussian
fits. Number in parentheses are 1$\sigma$ uncertainties in units
of the last digits. The observed positions are: TMC-1
$\alpha_{2000.0}$=04$^{\rm h}$41$^{\rm m}$41.9$^{\rm s}$,
$\delta_{2000.0}$=+25$^{\circ}$41$'$27.0$''$; Barnard 1
$\alpha_{2000.0}$=03$^{\rm h}$33$^{\rm m}$20.8$^{\rm s}$,
$\delta_{2000.0}$=+31$^{\circ}$07$'$34.0$''$; L134N
$\alpha_{2000.0}$=15$^{\rm h}$54$^{\rm m}$06.6$^{\rm s}$,
$\delta_{2000.0}$=-02$^{\circ}$52$'$19.1$''$; L1527
$\alpha_{2000.0}$=04$^{\rm h}$39$^{\rm m}$53.9$^{\rm s}$,
$\delta_{2000.0}$=+26$^{\circ}$03$'$11.0$''$; L483
$\alpha_{2000.0}$=18$^{\rm h}$17$^{\rm m}$29.8$^{\rm s}$,
$\delta_{2000.0}$=-04$^{\circ}$39$'$38.3$''$; Horsehead
$\alpha_{2000.0}$=05$^{\rm h}$40$^{\rm m}$53.7$^{\rm s}$,
$\delta_{2000.0}$=-02$^{\circ}$28$'$04.0$''$; Orion Bar
$\alpha_{2000.0}$=05$^{\rm h}$35$^{\rm m}$22.8$^{\rm s}$,
$\delta_{2000.0}$=-05$^{\circ}$25$'$01.0$''$; and NGC 7023
$\alpha_{2000.0}$=21$^{\rm h}$01$^{\rm m}$32.6$^{\rm s}$,
$\delta_{2000.0}$=+68$^{\circ}$10$'$27.0$''$.\\
$^a$ rms noise averaged over the line width. $^b$ 3$\sigma$ upper
limit assuming the same line width as the neutral. $^c$ $\eta_b$
is B$_{\rm eff}$/F$_{\rm eff}$, i.e. the ratio of T$_A^*$ to
T$_{mb}$. $^d$ Line observed with a spectral resolution of 80 kHz.
$^e$ The N=2-1 J=5/2-3/2 transition of CN at 226.8 GHz was
observed in this source with a S/N ratio higher than 6; line
widths are 0.58 km s$^{-1}$ in TMC-1, 1.15 km s$^{-1}$ in Barnard
1, 0.75 km s$^{-1}$ in L1527, 0.76 km s$^{-1}$ in L483, 0.70 km
s$^{-1}$ in the Horsehead, 1.76 km s$^{-1}$ in the Orion Bar, and
0.80 km s$^{-1}$ in NGC7023. $^f$ The N=3-2 transition of C$_2$H
at 262.0 GHz was observed in NGC 7023 with a S/N ratio higher than
20 and a line width of 0.65 km s$^{-1}$.}
\end{table*}

\section{Observations}

The observations were carried out with the IRAM-30m telescope from
24$^{th}$ July to 2$^{nd}$ August 2007. We searched for molecular
anions in the starless cores: TMC-1 (at the cyanopolyyne peak),
Barnard 1 (at the B1-b core) and L134N; and in the low mass
star-forming regions L1527 and L483, at the position of their
embedded IRAS sources. In PDRs, molecular anions are predicted to
have a peak abundance in the low extinction, $A_V$$\sim$1.5-3,
interface where the transition of C$^+$ to CO occurs
\citep{mil07}. We thus searched for anions toward such positions
in three PDRs: the Horsehead Nebula at the so-called IR peak where
the greatest hydrocarbon emission is observed \citep{tey04,pet05},
the Orion Bar at the (CO) position where CF$^+$ was discovered
\citep{neu06}, and the reflection nebula NGC 7023 at the so-called
PDR peak where CO$^+$ has been observed \citep{fue97}.

We used SIS receivers operating at 3 mm and 1 mm in
single-sideband mode with image rejections $>$ 20 dB at 3 mm and
$\geq$ 10 dB at 1 mm. System temperatures were 100-140 K at 3 mm
and 300-500 K at 1 mm. An autocorrelator was used as the back end.
The spectral resolution at 3 mm was 40 kHz ($\sim$ 0.13 km
s$^{-1}$), except for some observations of neutral species for
which 80 kHz was used (see Table~\ref{table-lineparameters}), and
80 kHz at 1 mm ($\sim$ 0.10 km s$^{-1}$). We used the frequency
switching technique with a frequency throw of 7.14 MHz. Pointing
and focus were checked every 1-2 hours observing nearby planets or
quasars. The list of observed molecules and transitions is given
in Table~\ref{table-lineparameters}.

\section{Results and discussion}

\begin{figure}
\includegraphics[angle=-90,scale=.48]{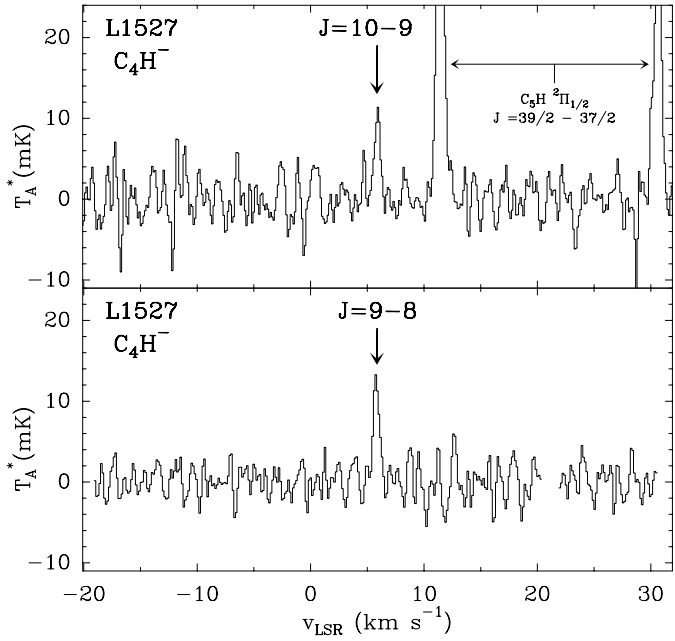}
\caption{J = 9-8 and J = 10-9 transitions of C$_4$H$^-$ observed
toward L1527 in 13.8 h and 22.1 h of integration time
respectively.} \label{fig-c4hm-lines} \vspace{-0.1cm}
\end{figure}

Most of the observing time was dedicated to the search for
C$_4$H$^-$ in TMC-1 and L1527, and for C$_6$H$^-$ in the Horsehead
Nebula. We could only detect C$_4$H$^-$ in L1527 through the J =
9-8 and J = 10-9 rotational lines (shown in
Fig.~\ref{fig-c4hm-lines} at a spectral resolution of 40 kHz)
which are observed with a signal-to-noise ratio in T$_A^*$ of 6
and 5 respectively. We thus confirm the tentative detection of the
J = 9-8 line announced by \citet{sak07a}. The lines have a
Gaussian-like profile with a width of 0.6 km s$^{-1}$ and are
centered at a source velocity of +5.85 km s$^{-1}$. Similar line
properties are also found for C$_4$H (see
Table~\ref{table-lineparameters}), for C$_6$H and C$_6$H$^-$
\citep{sak07a}, and for other carbon chains observed in L1527
\citep{sak07b}. The rms noise levels reached in the other searches
for anions, a few mK in T$_A^*$ in most of the cases, are given in
Table~\ref{table-lineparameters}.

We have calculated beam averaged column densities (see
Table~\ref{table-coldens}) under the LTE approximation. In L1527,
the column density of C$_4$H$^-$ is 1.6 $\times$ 10$^{10}$
cm$^{-2}$ and that of C$_4$H is 1.5 $\times$ 10$^{14}$ cm$^{-2}$.
The rotational temperature (T$_{\rm rot}$) obtained for C$_4$H is
14.1 $\pm$ 1.6 K, while in the case of C$_4$H$^-$ the two observed
lines suggests a value of 14 K which is in agreement with that
found for C$_4$H. In the rest of the cases T$_{\rm rot}$ has been
fixed: for the dense cores L1527 and L483 we assumed that T$_{\rm
rot}$ is equal to the gas kinetic temperature, 14 K and 10 K
respectively \citep{sak07b,taf00}, and adopted T$_{\rm rot}$ = 5 K
for the cold dark clouds and T$_{\rm rot}$ = 15 K for the PDRs.

The anion-to-neutral ratios derived in various sources from this
and previous works are given in Table~\ref{table-abunratios}. In
the C$_n$H$^-$ series, the ratios (or limits to ratios) observed
for C$_6$H$^-$ and C$_8$H$^-$ are always above 1 \%. They are
about two orders of magnitude lower for C$_4$H$^-$ and even
smaller for C$_2$H$^-$ (at least in IRC +10216). In the case of
the [CN$^-$]/[CN] ratio, the lowest upper limits derived (0.2 \%
for the lowest) are less constraining than in the case of
C$_2$H$^-$.

The anion-to-neutral ratio maybe expressed in terms of the
processes of formation and destruction of the anion. In steady
state we have (see \citealt{cer07}):
\begin{equation}
\frac{[\rm A^-]}{[\rm A]} = \frac{k_{\rm ea} [\rm e^-]}{k_{\rm B+}
[\rm B^+] + k_{\rm N^0} [{\rm N^0}] + \Gamma_{\rm ph} / \rm n}
\end{equation}
\noindent where n is the gas volume density and it is assumed that
the anion A$^-$ forms by attachment of an electron e$^-$ on A
(with a rate constant $k_{\rm ea}$) and can be destroyed by
reactions with cations B$^+$ ($k_{\rm B^+}$), neutral atoms N$^0$
($k_{\rm N^0}$), and by photodetachment ($\Gamma_{\rm ph}$).
Carbon chain anions are known to react rapidly with H and O atoms,
and somewhat more slowly with N atoms \citep{eic07}. Assuming
rather standard rate constants for the other destruction processes
($k_{\rm B^+}$ = 10$^{-7}$ cm$^3$ s$^{-1}$ and $\Gamma_{\rm ph}$
from \citealt{mil07}), we may use the observed [A$^-$]/[A] ratio
to derive the relevant electron attachment rate constant $k_{\rm
ea}$(A). This approach was used by \citet{cer07} in the context of
a chemical model of IRC +10216 to derive $k_{\rm ea}$(C$_4$H) and
$k_{\rm ea}$(C$_6$H). Here we have revised that model, including
loss of anions by reaction with atoms other than H, to arrive at:
\begin{equation}
\begin{array}{lcrccrr}
k_{\rm ea}({\rm C_8H}) & = & 2.5 & \times & 10^{-8}  & ({\rm
T}/300)^{-1/2} & {\rm cm^3 s^{-1}}\\
k_{\rm ea}({\rm C_6H}) & = & 1.4 & \times & 10^{-8}  & ({\rm
T}/300)^{-1/2} & {\rm cm^3 s^{-1}}\\
k_{\rm ea}({\rm C_4H}) & = & 9   & \times & 10^{-11} & ({\rm
T}/300)^{-1/2} & {\rm cm^3 s^{-1}}\\
k_{\rm ea}({\rm C_2H}) & < & 1.5 & \times & 10^{-11} & ({\rm
T}/300)^{-1/2} & {\rm cm^3 s^{-1}}
\end{array}
\end{equation}

The values of $k_{\rm ea}$(C$_4$H) and $k_{\rm ea}$(C$_6$H) at
temperatures of 10-30 K are similar to those derived in
\citet{cer07}. There exists a clear correlation between the value
of $k_{\rm ea}$ and the observed anion-to-neutral ratio, which
simply indicates that the latter is mostly controlled by the
electron attachment reaction, the destruction processes being
assumed similar for all anions. The rate constants $k_{\rm ea}$
obtained for C$_6$H and C$_8$H are of the same order of magnitude
as the theoretical values reported by \citet{mil07}, although the
value for C$_4$H is more than one order of magnitude lower than
the theoretical one. In fact, the chemical models of \citet{mil07}
systematically overestimate the anion-to-neutral ratio for
C$_4$H$^-$. They calculate values of the [C$_4$H$^-$]/[C$_4$H]
ratio in IRC +10216, in TMC-1 (at early time) and in the Horsehead
Nebula of 0.77 \%, 0.13 \% and 3.5 \% respectively, which are
higher by a factor of 20-100 than the values derived from
observations (see Table~\ref{table-abunratios}). It seems that the
theory of electron radiative attachment \citep{ter00} fails for
small molecules (the much lower dipole moment of C$_4$H compared
to C$_6$H and C$_8$H could play a role in diminishing the
efficiency for electron attachment; see \citealt{bru07b}), or that
there are important destruction processes for small anions so far
not considered (e.g. reaction with neutral molecules).

\begin{table}
\caption{Column densities} \label{table-coldens} \centering
\begin{tabular}{llrlrl@{ }r}
\hline \hline
\multicolumn{1}{c}{Source} & \multicolumn{2}{c}{Molecule} & \multicolumn{2}{c}{T$_{\rm rot}$(K)} & \multicolumn{2}{c}{N(10$^{10}$ cm$^{-2}$)} \\
\hline
TMC-1      & C$_4$H$^-$ & (C$_4$H) & 5.0$^a$  & (5.0$^a$)   & $<$3.7$^b$ & (71000) \\
           & CN$^-$     & (CN)     & 5.0$^a$  & (5.0$^a$)   & $<$140$^b$ & (4700) \\
           & C$_2$H$^-$ &          & 5.0$^a$  &             & $<$22$^b$  & \\
\hline
Barnard 1  & C$_4$H$^-$ & (C$_4$H) & 5.0$^a$  & (5.0$^a$)   & $<$6$^b$   & (25000) \\
           & CN$^-$     & (CN)     & 5.0$^a$  & (5.0$^a$)   & $<$84$^b$  & (21000) \\
           & C$_2$H$^-$ &          & 5.0$^a$  &             & $<$15$^b$ \\
\hline
L134N      & C$_4$H$^-$ & (C$_4$H) & 5.0$^a$  & (5.0$^a$)   & $<$3.8$^b$ & (6100) \\
           & CN$^-$     &          & 5.0$^a$  &             & $<$130$^b$ & \\
\hline
L1527      & C$_4$H$^-$ & (C$_4$H) & 13.6     & (14.1)      & 1.6        & (15000) \\
           & CN$^-$     & (CN)     & 14.0$^a$ & (14.0$^a$)  & $<$9.8$^b$ & ($>$4800) \\
           & C$_2$H$^-$ &          & 14.0$^a$ &             & $<$1.8$^b$ & \\
\hline
L483       & C$_4$H$^-$ & (C$_4$H) & 10.0$^a$ & (10.0$^a$)  & $<$2.3$^b$ & (6900) \\
           & CN$^-$     & (CN)     & 10.0$^a$ & (10.0$^a$)  & $<$90$^b$  & ($>$7100) \\
\hline
Horsehead  & C$_4$H$^-$ & (C$_4$H) & 15.0$^a$ & (15.0$^a$)  & $<$1.0$^b$ & (3000) \\
           & C$_6$H$^-$ & (C$_6$H) & 15.0$^a$ & (15.0$^a$)  & $<$8.0$^b$ & (90) \\
           & CN$^-$     & (CN)     & 15.0$^a$ & (15.0$^a$)  & $<$12$^b$  & (2200) \\
\hline
Orion Bar  & C$_4$H$^-$ & (C$_4$H) & 15.0$^a$ & (15.0$^a$)  & $<$1.6$^b$ & (2500) \\
           & CN$^-$     & (CN)     & 15.0$^a$ & (15.0$^a$)  & $<$46$^b$  & (27000) \\
\hline
NGC 7023   & CN$^-$     & (CN)     & 15.0$^a$ & (15.0$^a$)  & $<$37$^b$  & (1400) \\
           & C$_2$H$^-$ & (C$_2$H) & 15.0$^a$ & (15.0$^a$)  & $<$3.1$^b$ & (8900) \\
\hline
\end{tabular}
\tablenoteb{Parameters for the neutral species are given in
parentheses. $\mu$(C$_2$H$^-$) = 3.1 D; $\mu$(C$_4$H$^-$) = 6.2 D;
$\mu$(C$_6$H$^-$) = 8.2 D; $\mu$(CN$^-$) = 0.65 D. $^a$ Assumed
rotational temperature. $^b$ 3$\sigma$ upper limit.}
\end{table}

\begin{table*}
\caption{Anion-to-neutral ratios (in \%) in various molecular
sources} \label{table-abunratios} \centering
\begin{tabular}{lc@{}lc@{}l@{ }c@{}l@{}c@{}lc@{}l@{ }c@{}lc@{}l@{ }c@{}l@{}c@{}l}
\hline \hline
\multicolumn{1}{l}{}                           & \multicolumn{2}{c}{IRC +10216} & \multicolumn{2}{c}{TMC-1} & \multicolumn{2}{c}{Barnard 1} & \multicolumn{2}{c}{L134N} & \multicolumn{2}{c}{L1527} & \multicolumn{2}{c}{L483} & \multicolumn{2}{c}{Horsehead} & \multicolumn{2}{c}{Orion Bar} & \multicolumn{2}{c}{NGC 7023} \\
\hline
\multicolumn{1}{l}{[C$_2$H$^-$]/[C$_2$H].....} & $<$0.0030    & $^{(1,2)}$      & $<$0.033  & $^{(6,7)}$    &  $<$0.048    & $^{(6,7)}$     &  $<$0.25   & $^{(9,10)}$  & $<$0.0036 & $^{(6,7)}$    &              &           &              &                &              &                & $<$0.035     & $^{(6)}$      \\
\multicolumn{1}{l}{[C$_4$H$^-$]/[C$_4$H].....} & 0.024        & $^{(3)}$        & $<$0.0052 & $^{(6)}$      &  $<$0.024    & $^{(6)}$       &  $<$0.062  & $^{(6)}$     & 0.011     & $^{(6)}$      & $<$0.033     & $^{(6)}$  & $<$0.033     & $^{(6)}$       & $<$0.064     & $^{(6)}$       &              &               \\
\multicolumn{1}{l}{[C$_6$H$^-$]/[C$_6$H].....} & 6.2-8.6      & $^{(3,4)}$      & 1.6       & $^{(8)}$      &              &                &            &              & 9.3       & $^{(11)}$     &              &           & $<$8.9       & $^{(6)}$       &              &                &              &               \\
\multicolumn{1}{l}{[C$_8$H$^-$]/[C$_8$H].....} & 26           & $^{(5)}$        & 4.6       & $^{(8)}$      &              &                &            &              &           &               &              &           &              &                &              &                &              &               \\
\multicolumn{1}{l}{[CN$^-$]/[CN]........}      & $<$0.52      & $^{(1)}$        & $<$1.9    & $^{(6)}$      &  $<$0.40     & $^{(6)}$       & $<$6.5     & $^{(6,10)}$  & $<$0.20   & $^{(6)}$      & $<$1.3       & $^{(6)}$  & $<$0.55      & $^{(6)}$       & $<$0.17      & $^{(6)}$       & $<$2.6       & $^{(6)}$      \\
\hline
\end{tabular}
\tablerefs{(1) unpublished IRAM-30m data; (2) \citealt{cer00}; (3)
\citealt{cer07}; (4) \citealt{kas07}; (5) \citealt{rem07}; (6)
This work; (7) \citealt{sak07b}; (8) \citealt{bru07b}; (9)
\citealt{mor05}; (10) \citealt{dic00}; (11) \citealt{sak07a}.}
\end{table*}

The presence of molecular anions seems to be favored in IRC
+10216, where three different anions have been observed with
relatively high anion-to-neutral ratios, compared to the other
sources. The ratios are in particular larger in IRC +10216 than in
TMC-1, most probably due to a larger ionization degree in the
former.

Comparing the dense cores TMC-1 and L1527, we note that molecular
anions are present at a higher level in the latter, as indicated
by the larger [C$_6$H$^-$]/[C$_6$H] and [C$_4$H$^-$]/[C$_4$H]
ratios. This has been interpreted by \citet{sak07a} as due to a
higher gas density in L1527 ($\sim$ 10$^6$ cm$^{-3}$;
\citealt{sak07b}) compared to that in TMC-1 (10$^4$ cm$^{-3}$). An
increase in the gas density makes the abundance of H atoms
decrease more than that of electrons\footnote{It is predicted that
in a dense cloud the fractional abundance of H atoms varies with
the gas density as the inverse while that of electrons varies only
as the inverse of the square root (see e.g. \citealt{flo07}).}.
Thus according to Eq. 1 (the term $\Gamma_{\rm ph}$/n drops for a
dense cloud shielded from UV photons,) this results in an increase
of the anion-to-neutral-ratio which approaches its maximum value
$k_{\rm ea}$/$k_{\rm B^+}$, achieved in the limit when reactions
with cations dominate the destruction of anions, i.e. $k_{\rm
B^+}$ [{\rm B$^+$}] $\gg$ $k_{\rm N^0}$ [{\rm N$^0$}] and assuming
[{\rm B$^+$}] = [{\rm e$^-$}].

An interesting difference between TMC-1 and L1527 is that the
former is a starless quiescent core while the latter harbors an
embedded protostar (IRAS 04368+2557) with an associated bipolar
outflow and an infalling envelope of about 30'' \citep{oha97}. The
outflow is revealed in the high velocity ($\sim$ 3 km s$^{-1}$)
wings of some molecular lines such as HCO$^+$ J=1-0
\citep{sak07b}. These wings are not present in the C$_4$H$^-$
lines, which rules out the presence of a substantial fraction of
anion molecules in the outflow. The gravitational infall is
indicated by the two-peak asymmetry, with a brighter blue peak, in
the profiles of some optically thick lines of c-C$_3$H$_2$ and
H$_2$CO \citep{mye95}. Such a profile is not visible in any of the
C$_4$H and C$_4$H$^-$ lines observed in L1527, although they are
optically thin. Small mapping observations of the N=9-8 transition
indicates that C$_4$H has an extended distribution ($\sim$ 40'')
in all directions around the protostar and that the line width
increases when approaching to the central position \citep{sak07b}.
This latter signature was also observed by \citet{mye95} in the
c-C$_3$H$_2$ 2$_{1,2}$-1$_{0,1}$ line and was interpreted as due
to infall motion. Unfortunately, the rather weak C$_4$H$^-$ lines
observed in L1527 leave little chance to map their emission.

In the PDRs, the upper limits on the anion-to-neutral ratios are
not as low as in the rest of the sources due to the lower
abundance of carbon chains. Nevertheless, we find values much
lower than those predicted by chemical models. In particular,
\citet{mil07} overestimate by more than a factor of 100 the
anion-to-neutral ratio for C$_4$H$^-$ in the Horsehead Nebula. The
discrepancy could lie in the value of $k_{\rm ea}$(C$_4$H) used in
the models, which is probably too large. In the case of
C$_6$H$^-$, the predicted ratio (470 \%) is larger than our upper
limit by a factor of 50. The fact that chemical models reproduce
much better the observed [C$_6$H$^-$]/[C$_6$H] ratio in IRC +10216
and TMC-1 (within a factor of 4) than in the Horsehead Nebula
suggests that in PDRs the global destruction rate of the anion
could be underestimated.

This study has shown that molecular anions, despite the weakness
of their millimeter emission, may be used to probe the chemical
and physical conditions of interstellar clouds (see also
\citealt{flo07}). Further laboratory and theoretical work are
necessary for a more complete interpretation of the astronomical
observations.

While we were writing up this letter, we became aware of a paper
by \citet{sak07c} reporting the tentative detection of the
C$_4$H$^-$ J = 9-8 line in L1527. They derive a column density of
1.1 $\times$ 10$^{10}$ cm$^{-2}$ which is in good agreement with
the value obtained by us.

\begin{acknowledgements}
We would like to thank the IRAM staff and the 30m telescope
operators for their assistance during the observations and N.
Marcelino for useful advices on the frequency switching observing
mode. We also thank the anonymous referee for helpful suggestions.
We acknowledge N. Sakai, T. Sakai and S. Yamamoto for
communicating their observational results prior to publication.
This work has been supported by Spanish MEC trough grants
AYA2003-2785, AYA2006-14876 and ESP2004-665 and by Spanish CAM
under PRICIT project S-0505/ESP-0237 (ASTROCAM). MA also
acknowledges grant AP2003-4619 from Spanish MEC.
\end{acknowledgements}

\end{document}